\documentclass[prl,preprintnumbers,nofootinbib,aps,twocolumn]{revtex4-1}
\usepackage{epsfig,amsmath,amssymb}
\usepackage{mathrsfs}
\usepackage[normalem]{ulem}
\usepackage[usenames,dvipsnames]{color}
\usepackage[pagebackref=false, colorlinks=false]{hyperref}
\newcommand{\Lagr}{\mathcal{L}}
\newcommand{\mM}{\mathcal{M}}

\begin{document}

\title{Gravitational Waves with Orbital Angular Momentum}

\author{Pratyusava Baral}
\email{baralpratyusava@gmail.com} 
\affiliation{Department of Physics, Presidency University, Kolkata 700073, 
India.}

\author{Anarya Ray}
\email{ronanarya9988@gmail.com} 
\affiliation{Department of Physics, University of Winsconsin, Milwaukee  53211, United States.}

\author{Ratna Koley}
\email{ratna.physics@presiuniv.ac.in} 
\affiliation{Department of Physics, Presidency University, Kolkata 700073, 
India.}  

\author{Parthasarathi Majumdar}
\email{bhpartha@gmail.com}
\affiliation{School of Physical Sciences, Indian Association for the Cultivation of Science, Kolkata 700032, India.} 

\begin{abstract}

Compact orbiting binaries like the black hole binary system observed in GW150914 carry large amount of orbital angular momentum.  
The post-ringdown compact object formed after merger of such a binary configuration has only spin angular momentum, and this results in a 
large orbital angular momentum excess. One significant possibility is that the gravitational waves generated by the system 
carry away this excess orbital angular momentum. An estimate of this excess is made. Arguing that plane gravitational waves cannot possibly carry any orbital angular momentum, a case is made in this 
paper for gravitational wave beams carrying orbital angular momentum, akin to optical beams. Restricting to certain specific 
beam-configurations, we predict that such beams may produce a new type of strain, in addition to the longitudinal strains measured 
at aLIGO for GW150914 and GW170817. Current constraints on post-ringdown spins, derived within the plane-wave approximation of 
gravitational waves, therefore stand to improve.  The minimal modification that might be needed on a laser-interferometer detector 
(like aLIGO or VIRGO) to detect such additional strains is also briefly discussed.

\end{abstract}

\maketitle
\section{Introduction}
Gravitational waves (GWs) detected by the Advanced Laser Interferometer Gravitational Wave Observatory (aLIGO)  
\cite{LIGO1,LIGO2,LIGO3} have established the existence of inspiralling compact object binaries. Within General 
Relativity (GR), such systems radiate gravitational waves, carrying energy and angular momenta \cite{E1,E2}, 
while spiraling into each other. The amount of angular momentum carried, as viewed by an observer at infinity (assuming the space-time 
to be asymptotically flat) can be estimated by the difference in angular momentum of the initial and final stages of a merger. 
GW150914 confirmed the merger of two black holes separated by a radius of 210 km and of masses around $36\mM_\odot$ and $29\mM_\odot$, 
forming a resultant Kerr black hole of mass $\sim$ 62$\mM_\odot$ and a spin parameter of $\sim$ 0.67 \cite{LIGO1}.  
We estimate the rate of loss of orbital angular momentum by the coalescing binary, assuming the objects 
to be slowly moving, such that the quadrupole approximation to gravitational wave generation still holds. In this approximation, the rate of loss of 
orbital angular momentum can be given as 
\begin{eqnarray}
    \frac{dL^i}{dt}&=&\frac{c^3}{32\pi G} \epsilon^{ijk}\int d\Omega r^2 \langle \dot{h}_{ab}x^j \partial^k h_{ab} \rangle \nonumber \\
    &=&\frac{2G}{15c^5} \epsilon^{ijk} \langle \ddot{Q}_{ja}\dddot{Q}_{ka} \rangle \nonumber \\
    &\simeq& \frac{2G}{15c^5} (\mu r^2)^2 \omega^5 ~\label{estim}
\end{eqnarray} 
Inserting typical values for the masses ($\sim 30 M_{\odot}$) and the size of the compact binary ($\sim 200~km$), with the rotational angular frequency of about $10~Hz$, the rate of loss of orbital angular momentum from the system is about $10^{34}$ Joules - which is quite large. An actual estimate of orbital angular momentum radiated closer to merger would include higher order modes and thus increase the number. From such an estimate, the motivations for a serious analysis towards the possibility of {\it detection} of such a large orbital angular momentum, carried by the gravitational waves, in current and forthcoming laser interferometer experiments are very strong.

Measuring spacetime fluctuations as a function of time only, aLIGO has successfully constrained source parameters 
from which one can calculate radiated angular momentum \cite{Krev,N1,N2}. However a subtle issue arises if we want direct measurement of angular momentum at some future detector. As we show in the sequel, monochromatic plane waves with spatially constant polarizations, used predominantly in detection 
analysis, cannot carry orbital angular momentum. Specifically, gravitational waves cannot reach the detector as monochromatic 
plane waves due to angular momentum conservation. We stress that by the term `plane wave' we mean, a monochromatic wave solution of the linearized, source-free Einstein equation, not only whose constant phase surfaces are planes in spacetime, but also whose polarizations are {\it spatially constant}, as the wave propagates. Taking a clue from well established results in optics \cite{Allen}-\cite{o3}, we propose gravitational radiation with some phase structure or gravitational wave {\it beams} as a basis for expansion of the source waveform for detection analysis. The phase structure would enable us to measure orbital angular momentum directly. Recently, radiation from a spiralling charged particle \cite{Oangularmomentum1}-\cite{Oangularmomentum2} carrying orbital angular momentum in the context of classical electrodynamics, has been shown to possess a similar phase structure. 

There exist other motivational aspects of directly measuring angular momentum from gravitational waves. A direct measurement of orbital angular momentum would provide us with an estimate of its rate of loss from the inspiralling binary. This, in turn, might allow us to impose additional constraints on the parameters over and above those obtained by cross-correlation with various templates. This would further enable us to settle many controversies relating to various mergers \cite{NSC} like GW170817 \cite{LIGO3} (NS-NS merger) which are expected to be routine in the near future. Using this, we may also expect to put constraints on the exotic alternative compact objects like fuzz balls \cite{fb}, gravastars \cite{gs} \cite{req}, wormholes \cite{wh}, boson stars \cite{bs} and so on, and hence be able to ascertain better the actual composition of the coalescing binary. Comparing how well the estimated angular momentum loss of the system compares to the angular momentum carried by gravitational waves as detected by a faraway observer, additional restrictions on the validity of GR in the linearized regime may perhaps be ascertained. Lastly, the third-generation run of the aLIGO and VIRGO is expected to detect certain gravitational lensing events \cite{ref1}. An additional probe of angular momentum is expected to give us additional knowledge of the medium through which it passes. This may vastly improve our understanding of interactions of gravitational waves with matter as it passes through astrophysical objects such as stars or galactic clusters. However for the time-being, we focus on a direct independent study of orbital angular momentum carried by gravitational waves.

In this paper we examine the analytic structure of gravitational waves in the linearized regime, starting from the gauge fixed linearized vacuum Einstein equation. 
We further demand solutions that carry orbital angular momentum (We prove that plane waves do not carry orbital angular momentum in the next section). This naturally enables us to go beyond 
plane waves and discuss gravitational wave {\it beams}. However our solutions are well within the regime of General Relativity and does not take any modified 
gravity effect into account. We show {\it en passant} that plane waves cannot carry  orbital angular momentum, implying that recourse to gravitational wave 
beams is imperative. We choose a particular set of linearly independent beams which form a basis for gravitational waves with orbital angular 
momentum. Any gravitational radiation generated by sources within general relativity can be expanded in this beam basis. A brief discussion is presented on 
the effects these beams would have on spacetime. We also give a schematic outline, how these beams carrying orbital angular momentum may be 
detected and the contribution of the beam to the overall signal measured in a generic Laser-interferometer gravitational wave detector.

\section{Gravitational Wave beams}

We employ here the linearized {\it tetrad} formalism \cite{tetrad} for discussing gravitational waves, for two reasons : the ease to discuss fermionic interactions in astrophysically relevant quantum field theoretic analysis, and to understand better the transition from local Lorentz invariance to global Lorentz symmetry under linearization - a phenomenon which remains slightly obscure within the metric formalism. However, the prescription to change to metric computations is included, for the ease of the readers.

For the purpose of linearization, the spacetime tetrad components $e^{\mu}_a$ are decomposed as $e_{\mu a}=\widehat{e}_{\mu a} + 
\varepsilon_{\mu a}$, where $\widehat{e}_{\mu a}$ is the background Minkowski spacetime tetrad and $\varepsilon_{\mu a}$ is the linear 
fluctuation. Linearized gravity in the harmonic gauge using this perturbed tetrad can be expressed as, 
$\widehat{e}^a_\mu \square^2 \varepsilon_{a \nu} + {\widehat{e}^a}_\nu \square^2 \varepsilon_{a\mu}= 0$ where Greek indices specify  
spacetime labels and early Latin indices are tangent space labels. The late Latin indices are reserved for three dimensional space. Therefore, the metric and fluctuations turns out to be 
\begin{eqnarray}
&~&\eta_{\mu \nu}= \eta_{ab}{\widehat{e}^a}_\mu {\widehat{e}^b}_\nu \\
&~& h_{\mu \nu}={\widehat{e}^a}_\mu \varepsilon_{a\nu}+{\widehat{e}^a}_\nu \varepsilon_{a\mu}+\textbf{O}(\varepsilon^2) \label{eq:trans}
\end{eqnarray}
Eq. \eqref{eq:trans} explicitly states how to transform from the metric fluctuations to the tetrad fluctuations and vice-versa.

The gauge fixed linearized tetrad equation admits a wave like solution given by,
\begin{equation}\label{eq:sub}
\varepsilon_{a\mu}=\vartheta_{a\mu}(x^\sigma)~ \exp(ik_\lambda x^\lambda) + \vartheta^{*}_{a\mu}(x^\sigma)~ \exp(-ik_\lambda x^\lambda)
\end{equation}
with the asterisk ($^*$) representing complex conjugation. This solution imposed on linearized tetrad equation gives,
\begin{equation}\label{eq:HE1}
(\partial^\nu \partial_\nu + 2ik_\nu\partial^\nu)~\vartheta_{a\mu}(x) = 0~.
\end{equation}
The Lagrangian density and the energy-momentum tensor for linearized gravity \cite{LL} are 
\begin{eqnarray}
\Lagr
&=&-\frac{c^4}{32\pi G}(\partial^\mu \varepsilon_{a\sigma} \partial_\mu \varepsilon^{a \sigma}+{\widehat{e}^a}_\sigma \widehat{e}^{b\rho}\partial^\mu \varepsilon_{a\rho} \partial_\mu {\varepsilon_b}^\sigma) ~\label{lagr}\\
T^{\mu\nu}&=&
\frac{c^4}{16\pi G}(\partial^\mu \varepsilon_{a\sigma} \partial^\nu \varepsilon^{a \sigma}+{\widehat{e}^a}_\sigma \widehat{e}^{b\rho}\partial^\mu  \varepsilon_{a\rho} \partial^\nu {\varepsilon_b}^\sigma)
\end{eqnarray}

Since we are dealing with small fluctuations around a Minkowski spacetime tetrad in the linearized region, the system is globally Lorentz-symmetric. The conserved Noether charge density corresponding to this symmetry can be expressed as,
\begin{eqnarray}
\frac{8\pi G}{c^2} {\cal M}_{\rho \sigma} = \dot{\varepsilon}_{a\mu}
[x_{[\rho|}(\partial_{|\sigma]}\varepsilon^{a\mu}+{\widehat{e}^a}_\nu \widehat{e}^{b\mu}\partial_{|\sigma]}{\varepsilon_b}^\nu)] \label{eq:angul}
\end{eqnarray}
where, over dot represents time derivative, and the square brackets denote anti-symmetry. Integrating this charge density over all space gives the infinitesimal Lorentz generators. 

If $\vartheta_{a\mu}$ is not a function of the spatial coordinates ($x^i $) on integration the first term reduces to terms like $\int (x^ik^j-x^j k^i)d^3 x$; with the integration measure being clearly rotationally invariant, the first term {\it vanishes by spatial rotational invariance !} This implies that a spatially constant polarization, like for plane waves in our connotation (which definitely satisfy equation \eqref{eq:HE1}), cannot carry orbital angular momentum. It follows that for gravitational waves to carry orbital angular momentum, their polarization tensors must themselves be tensor {\it fields}. 
One way to enable polarization fields in gravitational waves is through gravitational wave beams, akin to optical beams. Plane waves definitely carry spin angular 
momentum. But the spin angular momentum density is miniscule, and definitely cannot account for all angular momentum radiated. Moreover the spin part is a 
property of the radiation and can never be zero for a rank two tensor field. The orbital angular momentum is dependent on orientation of objects and fields and 
thus somewhat arbitrary. Thus there is no reason to expect that  orbital angular momentum of source is getting converted to spin angular momentum of the wave. 
Thus we are looking for the most general solution of linearized vacuum Einstein field equations.

\subsection{Laguerre-Gaussian (LG) Beams}

Let $\textbf{z}$ direction be the direction of propagation of the gravitational wave beam. Since any conceivable detector has to be placed far away from the sources, 
we can safely assume the beam to be paraxial ($k_z \approx k$) \cite{o2, o2-1}. So equation \ref{eq:HE1} can be expressed as
\begin{equation}\label{eq:Hel}
({\nabla_T}^2 + \partial^z\partial_z -\partial^t\partial_t+ 2ik_z\partial^z - 2ik_t\partial^t)~\vartheta_{a\mu }(x) = 0
\end{equation}
where ${\nabla_T}^2 $ is a two dimensional Laplace operator in the plane perpendicular to $\textbf{z}$. The paraxial approximation also guarantees that the 
change in polarization tensor in the direction of propagation is negligible in comparison to the wave vector 
$\left( \left| \frac{\partial ^ 2 \vartheta^{a\mu}}{\partial z^2} \right|~ \ll ~ \left| k_z \frac{\partial \vartheta^{a\mu }}{\partial z}\right| \right)$. 
So equation (\ref{eq:Hel}) reduces to, $({\nabla_T}^2 + 2ik_z\partial^z)~\vartheta_{a\mu}(x) = 0$.
We choose the transverse plane to be spanned by $(r,\phi)$ then $\nabla_T^2$ = $\frac{1}{r}\partial_r+{\partial_r}^2 + \frac{1}{r^2}{\partial_\phi}^2$.

For simplicity, we choose to work with one particular component of $\vartheta_{a\mu}$ and for the time being we drop the spacetime and tangent space indices.

A solution of the form 
\begin{widetext}
\begin{eqnarray}\label{eq:LGB}
\vartheta_{mp}(r,\phi,z) = \frac{A_{mp}}{w(z)}~\left(\frac{\sqrt{2}r}{w(z)}\right)^{|m|}  
\exp\left[\frac{- i k r^2 z}{2(z^2+z^2_R)}\right] 
L^{|m|}_p\left(\frac{2r^2}{w^2(z)}\right)  && \exp\left[im\phi - i(2p+|m|+1)~tan^{-1}\frac{z}{z_R}\right] \nonumber \\
&& \times \exp\left(\frac{-r^2}{w^2(z)}\right)
\end{eqnarray}
\end{widetext}
satisfies equation (\ref{eq:Hel}) in its paraxial form \cite{o2-1} with $m,p$ taking integer values referring to various modes. The radius of the beam $w(z)$ is given by $w(z)=w(0)\sqrt{1 + \left( \frac{z}{z_R}\right) ^2 }$ where $z_R=\frac{\pi w(0)}{\lambda}$ and $k \approx k_z  = \frac{2\pi}{\lambda}$. $L^{|m|}_p(x)$ is the associated Laguerre polynomial while $A_{mp}$ is a normalization constant. By definition of Laguerre polynomials, $p$ has to be an integer. The single valuedness of the field under a rotation of $\pi$ radian forces the azimuthally dependent phase factor to be quantized with $m$ taking only integral values. 
The Laguerre-Gaussian modes are orthonormal in both labels $m$ \& $p$ such that
$$\int_0^{2\pi} d\phi ~ \int _0 ^\infty r dr ~\vartheta_{mp}(r,\phi,z)[\vartheta_{qr}(r,\phi,z)]^*= \delta_{mq}\delta_{pr}.$$
This guarantees that the LG modes form a complete orthonormal family which can be used as a basis for a beam 
with an arbitrary polarization distribution. These beams exhibit a symmetry manifest in the use of cylindrical coordinates. Any other choice of a complete set of beams a different symmetry (like, e.g., the Bessel beam) is equally valid; these beams can of course be expressed as a linear combination of LG modes. Since our solutions form a basis, any waveform including plane waves or spherical waves can be expanded in our basis. 

The orbital angular momentum density can be expressed in a simple form which follows directly from equation \eqref{eq:angul}, 
\begin{eqnarray}
      \vec{L} &=&\int d^{3}x \{ \left [-\frac{l}{\omega} \frac{z}{r}|\vartheta_{mp}|^{2}\hat{r}+\frac{r}{c}\left(\frac{z^{2}}{z^{2}+z_{R}^{2}}-1\right)|\vartheta_{mp}|^{2}\hat{\phi} \right] \nonumber \\
      &+& \frac{l}{\omega}|\vartheta_{mp}|^{2}\hat{z} \}\label{Oangular momentum}
\end{eqnarray}
Such solutions also exist in the case of electromagnetic waves and have been extensively studied in refs. \cite{Allen}-\cite{o3}. Bialynicki-Birula et. al. \cite{BB} have also discussed GW beams using an electromagnetic-gravitational correspondence within a spinoral formalism. In this work we take an approach that appears to be more convenient for phenomenological applications.

Integrating the angular momentum density \eqref{Oangular momentum} over all space, we get the total angular momentum. As stated earlier our analysis 
is only valid in the weak field regime. So, the integration domain must be restricted to that regime. Here we 
should mention the caveat that unlike in laser optics, $w(0)$ has no significance and is just a 
free parameter of the chosen basis.

\section{Effect of a Passing Gravitational Wave Beam on Spacetime : Possible Detection}

We first describe the effect on spacetime in the plane perpendicular to the direction of propagation. Let this transverse plane be spanned by coordinates $x,y$. It is well known that in standard TT gauge all perturbation except $h_+~=~{\widehat{e}^a}_{x}{\varepsilon^b}_{x}+{\widehat{e}^b}_{x}{\varepsilon^a}_{x}$ and $h_\times~=~{\widehat{e}^a}_{x}{\varepsilon^b}_{y}+{\widehat{e}^b}_{y}{\varepsilon^a}_{x}$ can be gauged away to $0$. To start with let us consider a  gravitational wave beam consisting of only the component $h_+$ . The corresponding tetrad fluctuation $\varepsilon_{a\mu=x}$ contains an infinite number of various LG modes. The proper distance between four test particles localized at $A(0,0), B(L,0), C(L,L)$ and $D(0,L)$ would change due to the passing gravitational wave. Firstly let us consider motion along or parallel to $x$-axis. Therefore, $dt^2~=~dy^2~=~dz^2~=0$.
\begin{eqnarray}
ds^2 &=& (1+2{\widehat{e}^a}_{x} \varepsilon_{a x})dx^2 \nonumber \\
\Rightarrow \Delta l &\simeq& \int_0^l \widehat{e}^a_x \varepsilon_{ax}dx \simeq f(y) \label{shr}\\
\Rightarrow \Delta l &=& \int_0^l (C+g(y))dx\\
&=& Cl+\int_0^l g(y)dx \label{eq:NLS}
\end{eqnarray}
where, $f(y)$ and $g(y)$ are functions of $y$ differing with a constant value of C.

The decomposition of the integral into two parts, one involving a constant C which is the dominant plane wave contribution; the sub-dominant contribution is due to the beam structure. The first term gives normal longitudinal strain as expected by plane waves and the second is a non-longitudinal term. The second term which is essentially due to deviation from plane waves carry all information about radiated angular momentum giving rise to various new non-trivial effects.

If a gravitational wave of constant polarization passes over a circular ring of particles (shown by bold green dots), they would change to an elliptical ring (shown by red dots) as shown in Fig.1. 
\begin{figure}[htb]
\centering
 \includegraphics[width=7cm,height=7cm]{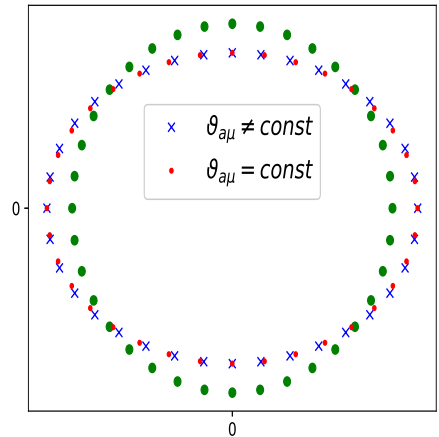}
 \label{Fig.1}
\caption {A ring of masses in presence and absence of GWs}
\end{figure}
The presence of a LG mode would deviate the masses from their expected places (shown by blue crosses) due to the second term in equation \eqref{eq:NLS}. The deviation from an ellipse for the lowest mode is plotted in Fig.2. The symmetry in the figure is precisely due to the symmetry in $x$ and $y$ coordinates for $m=0$ mode. Since each $h_+$ and $h_\times$ will contain an infinite number of LG modes, both these polarizations will contain a smear of infinitely many polarization states.
\begin{figure}[htb]
\centering
 \includegraphics[width=6cm,height=4.5cm]{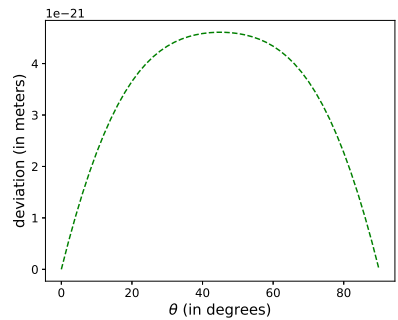}
 \label{Fig.2}
\caption {Absolute deviation from an ellipse}
\end{figure}

\subsection{Effect of Gravitational Wave beams on Laser-interferometer detectors}

Observational evidence for gravitational wave beams necessitates direct measurement of the polarization as a function of space and time. Unfortunately it is not easy to achieve this for extant laser interferometer set-ups. The important aspect is that the change in the length of an interferometer arm is no longer a simple linear function of its original length, if the interferometer is exposed to a gravitational wave beam. This departure from linearity can be exploited to infer information about possible orbital angular momentum of the incident gravitational wave. 

We shall now attempt to assess the effect of gravitational wave beams with the stated polarization profile on laser interferometer detectors. Let the plane of the detector be spanned by $x,y$ coordinates and the direction of propagation make an arbitrary angle with the $z$-axis. This detector will measure the intensity of the component of the incoming beam of gravitational waves along the $z$ direction.

For laser beams travelling along the arms of a detector in the $xy$  plane, $ds^{2}=c^2dt^{2}-dx^{2}\{1+({\widehat{e}^a}_{x}{\varepsilon^b}_{x}+{\widehat{e}^b}_{x}{\varepsilon^a}_{x})\eta_{ab}\}-dy^{2}\{1-({\widehat{e}^a}_{x}{\varepsilon^b}_{x}+{\widehat{e}^b}_{x}{\varepsilon^a}_{x})\eta_{ab}\}-dxdy({\widehat{e}^a}_{x}{\varepsilon^b}_{y}+{\widehat{e}^b}_{y}{\varepsilon^a}_{x})\eta_{ab} = 0$.
Now for the $x$-arm (length=$L$), $y=0=dy$, and for the $y$-arm (length=$L$), $x=0=dx$. Thus,
\begin{equation}
\begin{aligned}
    & cdt_{x}=dx\sqrt{1+2{\widehat{e}^a}_{x}\varepsilon_{ax}|_{y=0}} \approx dx(1+{\widehat{e}^a}_{x}\varepsilon_{ax}|_{y=0}) & \\
\end{aligned}
\end{equation}

Using equation (\ref{eq:sub}) and putting t=$\frac{x}{c}$ or t=$\frac{y}{c}$ for the forward journey 
and t=$\frac{L+x}{c}$ and $\frac{L+y}{c}$ for the return journey, \cite{Schutz} we get: $\tau_{x}=2L+\int^{L}_{0} \widehat{e}^{a}_{x}(\vartheta_{ax}(x,0,z)e^{ik(z-x)}+\vartheta^{*}_{ax}(x,0,z)e^{-ik(z-x)})dx +\int_{0}^{L} \widehat{e}^{a}_{x}(\vartheta_{ax}(x,0,z)e^{ik(z-(L+x))}+\vartheta^{*}_{ax}(x,0,z)e^{-ik(z-(L+x))})dx$.

Similar equations hold for the $y$-direction. Now we can calculate the phase difference between the $x$ and $y$ arm light rays:
\begin{equation}
    \delta \phi(L) =\frac{2\pi(\tau_{x}-\tau_{y})}{\lambda_{light}}=f(L)\approx \alpha+\beta L+\gamma L^2 \label{beam}
\end{equation}
where $\alpha$,$\beta$ and $\gamma$ are constants. For plane waves $\gamma=0$.

Figure (3) shows how $\frac{\delta \phi(L)}{\delta \phi(4 km)}$ depends on $L$ when $m = p = 0$. It clearly deviates from the linear nature shown by the blue dots. The value of 4 km is chosen for our numerical analysis. It has no special significance and any other choice is equally valid. The ratio does not depend on the value of $w(0)$ (in the range $10^2 - 10^8$m) in this case. For $m = 1$, the nature of the plot remains similar to that of figure (3), although it is now sensitive to $w(0)$.  The strains definitely depend on the value of $w(0)$ (as well as $L$) as shown in figure (3). As is clear from the graph, typical strain values predicted for a binary black hole merger source as reported for GW150914, is about $\sim 10^{-21}$ which is very much detectable by aLIGO type detectors. If $m=1$ we have a $(1/z)$ factor suppression of the gravitational wave signal, leading to additional strain (due to orbital angular momentum carried by gravitational waves) values that are far smaller than current sensitivities.
\begin{figure}[h]
\centering
 \includegraphics[width=6cm,height=4.5cm]{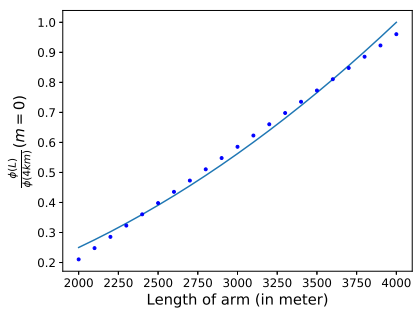}
 \label{Fig.3}
\caption {The variation of $\frac{\delta \phi(L)}{\delta \phi(4 km)}$ with interferometer arm length has been shown for  $m=0, p=0$.}
\end{figure}

\subsection{Detection Scheme}

Thus, it can be seen that the unlike the case for a plane wave where $\delta \phi$ is directly proportional to the wave amplitude and 
the arm length $L$, if the incident wave carries angular momentum and hence a beam of polarization, the phase difference will vary as a nonlinear and complicated function of arm-length.

If it is possible to vary the arm length, we can measure $\delta \phi$ for different values of $L$, and by comparing the data obtained 
with the functional dependence shown in figures (3) and using sophisticated statistical techniques and proper source modeling, 
one may directly measure the total angular momentum carried by gravitational waves. 

\section{Conclusions}

As electromagnetic beams carry orbital angular momentum, we have shown in this paper that there is sufficient reason to expect the same for gravitational waves. However, the main point of difference is that unlike lasers we cannot make any form of `gaser' (a gravitational laser!!) and thus are dependent on nature to produce gravitational wave beams that carry orbital angular momentum. Luckily it turns out that the simplest detectable gravitational wave emitting system, should radiate angular momentum. Unlike man-made lasers which usually have 1 particular mode, a gravitational wave will be a collection of various modes. Although any particular mode of a specific class of beams, can be expressed as a linear superposition of various modes of another class of beams, we have chosen LG modes as they are perhaps the simplest and most elegant solution, suffering no problem of divergence at asymptotic regions. 

In this paper, in addition to showing the necessity of considering gravitational wave beams in place of plane waves in order to explain the orbital angular momentum emitted via gravitational waves, by the merger of inspiralling compact binaries, we have presented an account of the effects these gravitational wave beams will have on spacetime in general and on laser interferometer detectors. Further, perhaps for the first time, we have proposed a schematic way of measuring the phase structure of gravitational radiation, by incorporating minimal changes in extant interferometers. Since the orbital angular momentum of gravitational waves can be directly calculated from these amplitudes, we have thus, again for the first time, proposed a schematic method of direct measurement of angular momentum carried by gravitational waves.

We are primarily interested in the lowest order mode, because the first higher mode for GW150914 like sources will have non-unique values of strains dependent on the normalization factor $w(0)$ and primarily because we have a $1/z^m$-suppression. This might make the interference signal too weak to detect. Having said that though, the expressions are dependent on various non-linear parameters and a particular set of $w(0)$; tweaking the frequency and distance it may be possible to produce such additional strains for higher modes, which are more realistic.

The idea of gravitational wave beams is at its infancy and thus lot of work remains to be done. From the values of the plus and cross polarizations at asymptotic regions obtained from various numerical simulations one might try to estimate various beam parameters. \textcolor{black}{ Since the beams form a complete basis, any wave form predicted by numerical relativity can always be expanded in terms of the beams by computing the overlap integrals and the corresponding beam parameters measured via the scheme discussed in this paper}. This would enable us to understand if data from a fixed length detector is sufficient to determine the beam parameters uniquely with a little degeneracy. Although this work is yet to be done, our intuition suggests that the deviation from plane waves will not become apparent due to their smallness until we have some way to vary the length of detectors. However a more rigorous calculation is left to be done before making any concrete statement.

Moreover the solution presented here involves the paraxial approximation. It is possible to get exact solutions of the vacuum linearized Einstein's equations beyond the paraxial approximation, which still carry orbital angular momentum and form a complete basis. This is currently under study.

\section{Acknowledgements}

The authors would like to thank Soumendra Kishore Roy and Sk. Sajid of Presidency University, Kolkata for valuable discussions and inputs. R. Koley 
acknowledges WBDHESTBT research fund. PB and AR would like to thank Bala Iyer of ICTS, K. G. Arun and J. Haque of CMI,India and A. Gupta of Pennsylvania State University for discussions in the IAGRG meeting 2018. PB is grateful to Luc Blanchet for a discussion at the 2019 ICTS Summer School on Gravitational Waves regarding total orbital angular momentum  radiated away through gravitational waves.


\begin{thebibliography}{99}
\bibitem{LIGO1}B. P. Abbott et al., Phys. Rev. Lett. 116 (2016) 241102, \href{https://arxiv.org/abs/1602.03840}{arXiv:1610.02182} [gr-qc].
\bibitem{LIGO2}B. P. Abbott et al., Phys. Rev. Lett.118 (2017) 221101, \href{https://arxiv.org/abs/1706.01812}{arXiv:1706.01812} [gr-qc].
\bibitem{LIGO3}B. P. Abbott et al., Phys. Rev. Lett. 119 (2017) 161101, \href{https://arxiv.org/abs/1710.05832}{arXiv:1710.05832} [gr-qc]. 
\bibitem{E1}
A. Einstein, Sitzungsber. K. Preuss. Akad. Wiss.1, 688(1916).
\bibitem{E2}
A. Einstein, Sitzungsber. K. Preuss. Akad. Wiss 1, 154(1918).
\bibitem{refchan}
S. Stevenson et al., Nat Commun. 10.1038/ncomms14906(2017), \href{https://arxiv.org/abs/1704.01352}{arXiv:1704.01352} [astro-ph.HE].
\bibitem{ref2}
I. Mandel, A. Farmer, \href{https://arxiv.org/abs/1806.05820}{arXiv:1806.05820} [astro-ph.HE].
\bibitem{BB2}
I. Bialynicki-Birula, S. Charzyński, Phys. Rev. Lett. 121 (2018) 171101 \href{https://arxiv.org/abs/1810.02219}{arXiv:1810.02219} [gr-qc]. 
\bibitem{Krev}
K. Thorne, Rev. Mod. Phys. 52, 299 (1980).
\bibitem{N1}
T. W. Baumgarte, S. L. Shapiro, Numerical Relativity: Solving Einstein's Equations on the Computer 1st Edition (Cambridge University Press; 1 August 16, 2010)
\bibitem{N2}
M. Shibata, Numerical Relativity (100 Years of General Relativity) (World Scientific Publishing Company November 5, 2015)
\bibitem{Allen}
L. Allen et. al., Phys. Rev. A, vol. 45, no. 11, pp. 8185-90, June 1992, reprinted in [57,Paper 2.1]. 
\bibitem{o1}
A. M. Yao, M. J. Padgett, Adv. Opt. Phot. 3 161 (2011).
\bibitem{o2}
J. B. Gotte, S. M. Barnett, Light beams carrying orbital angular momentum, Chapter 1 of THE ANGULAR MOMENTUM OF LIGHT edited by D. Andrews and M. Babiker, Cambridge University Press, 1 edition (2013). 
\bibitem{o2-1}
H. A. Haus, Waves and Fields in Optoelectronics (Chapter 5), Prentice-Hall Inc. (1984)
\bibitem{o3}
D. S. Simon, A guided tour of light beams-Morgan $\&$ Claypool Publishers(IOP concise physics) (2016).
\bibitem{Oangularmomentum1} M Katoh et al., Phys. Rev. Lett. 118 (2017), 094801, \href{https://arxiv.org/abs/1610.02182}{ arXiv:1610.02182} [physics.optics].
\bibitem{Oangularmomentum2} V. Epp, U. Guselnikova, Physics Letters A Volume 383(2019), Issue 22.
\bibitem{ref1}
Christian et al., Phys. Rev. D 10 (2018) 103022 \href{https://arxiv.org/abs/1802.02586}{arXiv:1802.02586} [astro-ph.HE].
\bibitem{NSC}
H. Yang et. al., \href{https://arxiv.org/abs/1710.05891}{arXiv:1710.05891} [gr-qc].
\bibitem{fb}
S. D. Mathur, Fortsch. Phys. 53 (2005)793–827, \href{https://arxiv.org/abs/hep-th/0502050}{arXiv:hep-th/0502050} [hep-th].
\bibitem{gs}
P. O. Mazur, E. Mottola, \href{https://arxiv.org/abs/gr-qc/0109035}{arXiv:gr-qc/109035} [gr-qc].
\bibitem{req}
C. Chirenti, L. Rezzolla, Phys. Rev. D94 (2016) 084016, \href{https://arxiv.org/abs/1602.08759}{arXiv:1602.08759} [gr-qc].
\bibitem{wh}
T. Damour, S. N. Solodukhin, Phys. Rev. D76 (2007) 024016, \href{https://arxiv.org/abs/0704.2667}{arXiv:0704.2667} [gr-qc].
\bibitem{bs}
R. Ruffini, S. Bonazzola, Phys. Rev. 187 (1969) 1767–1783.
\bibitem{ECO}
Z. Mark et al., Phys. Rev. D 96 (2017) 084002, \href{https://arxiv.org/abs/1706.06155}{arXiv:1706.06155} [gr-qc].
\bibitem{tetrad}
J. Yepez, \href{https://arxiv.org/abs/1106.2037}{arXiv:1106.2037} [gr-qc].
\bibitem{LL}
M. Maggiore,
Gravitational Waves. Vol. 1: Theory and Experiments, Oxford Master Series in Physics (Oxford University Press, 2007).
\bibitem{BB}
I. Bialynicki-Birula, Z. Bialynicka-Birula, New J. Phys. 18 (2016) 023022, \href{https://arxiv.org/abs/1511.08909}{arXiv:1511.08909v1} [gr-qc].
\bibitem{Schutz}
B. Schutz, A First Course in General Relativity, Cambridge University Press; 2nd edition (June 22, 2009).
\end{thebibliography}
\end{document}